%% file: EGauthorGuidelines-Rendering2021-sub.tex
\documentclass{egpubl}
\usepackage{egsr2021}
 
\SpecialIssueSubmission    

\usepackage[T1]{fontenc}
\usepackage{dfadobe} 

\usepackage{booktabs}
\usepackage{threeparttable}
\usepackage{multirow}
\usepackage{amstext}
\usepackage{ulem} 

\usepackage{cite}  
\BibtexOrBiblatex
\electronicVersion
\PrintedOrElectronic
\ifpdf \usepackage[pdftex]{graphicx} \pdfcompresslevel=9
\else \usepackage[dvips]{graphicx} \fi

\usepackage{egweblnk} 

\newcommand{\ren}[1]{\textcolor[rgb]{0.5,0.5,1.0}{Haocheng: #1}}
\newcommand{\fan}[1]{\textcolor[rgb]{1.0, 0.6571, 0.0}{Fan: #1}}
\newcommand{\revisionFan}[1]{#1}
\newcommand{\huo}[1] {
  \textcolor{cyan}{\bfseries{Huo: {#1}}}
}
\newcommand{\Skip}[1] {}

\title[Real-time MC Denoising with Weight Sharing Kernel Prediction Network]{Real-time Monte Carlo Denoising with Weight Sharing Kernel Prediction Network}

\author[]
{\parbox{\textwidth}{\centering 
Hangming Fan, Rui Wang, Yuchi Huo, Hujun Bao\\
State Key Lab of CAD\&CG, Zhejiang University}
}

\begin{document}


\maketitle

\begin{abstract}

Real-time Monte Carlo denoising aims at removing severe noise under low samples per pixel (spp) in a strict time budget. Recently, kernel-prediction methods use a neural network to predict each pixel's filtering kernel and have shown a great potential to remove Monte Carlo noise. However, the heavy computation overhead blocks these methods from real-time applications. This paper expands the kernel-prediction method and proposes a novel approach to denoise very low spp (e.g., 1-spp) Monte Carlo path traced images at real-time frame rates. Instead of using the neural network to directly predict the kernel map, i.e., the complete weights of each per-pixel filtering kernel, we predict an encoding of the kernel map, followed by a high-efficiency decoder with unfolding operations for a high-quality reconstruction of the filtering kernels . The kernel map encoding yields a compact single-channel representation of the kernel map, which can significantly reduce the kernel-prediction network's throughput. In addition, we adopt a scalable kernel fusion module to improve denoising quality. The proposed approach preserves kernel prediction methods' denoising quality while roughly halving its denoising time for 1-spp noisy inputs. In addition, compared with the recent neural bilateral grid-based real-time denoiser, our approach benefits from the high parallelism of kernel-based reconstruction and produces better denoising results at equal time.


\end{abstract}  


\section{Introduction}

Monte Carlo path tracing is a technique used to synthesize realistic images, and it has the attribute of using a unified rendering pipeline to simulate various physically-based visual effects, which is appealing to both offline and real-time applications. But the use of Monte Carlo integration requires thousands of samples per pixel to produce a visually noise-free path traced image, which leads to significant computational cost. Given the computation budget of desktop GPU in the near future, only one 
\revisionFan{sample per pixel (spp)}
is allowed in real-time path tracing application \cite{10.1145/3072959.3073601,wang2013gpu,huo2015matrix,huo2016adaptive,huo2020spherical}. While signal simplification is common in some low-quality rendering applications \cite{zhang2021powernet,li2020automatic,kim2020single,li2021multi,an2021hypergraph,park2021meshchain}, physically accuracy is important for interactive global illumination. Therefore, using image-based denoising techniques to remove Monte Carlo noise would be a practical solution for production-ready applications.

Typically the low-spp Monte Carlo path traced image provides limited lighting information for denoising, and it is challenging to reconstruct unbiased results. Most denoising techniques introduce bias to reduce the variance, such as gathering the information from nearby pixels to reconstruct the center pixel color. Besides, the noise-free auxiliary feature buffers such as shading normals and texture albedo colors can provide geometric and material information about the scene, which can be used as inputs to guide the denoising algorithm. These buffers can be obtained in the path tracing phase with a negligible time cost. 

Given the success of convolutional neural network (CNN) in solving computer vision and graphics tasks, lots of recent 
\revisionFan{researchers}
focus on using a CNN to build their denoising architecture. Bako et al. \cite{10.1145/3072959.3073708} proposed to use a neural network to predict per-pixel filtering kernels instead of using hand-crafted kernels. Vogels et al. \cite{10.1145/3197517.3201388} then extended this architecture to predict small kernels in hierarchical resolutions for saving computations. This multi-resolution denoising architecture was also used in the layer-based denoiser \cite{munkberg2020neural} and the temporal adaptive sampling method \cite{hasselgren2020neural}. 
These methods have shown the capability of the kernel prediction method in Monte Carlo denoising. However, they 
\revisionFan{mostly} 
target offline applications and rarely achieve 
\revisionFan{real-time} 
frame rates. The high computational cost of 
\revisionFan{predicting} 
large filtering kernels makes applying such techniques to real-time infeasible. 

In this paper, we propose a novel approach to reduce the kernel prediction methods' overhead and achieve real-time Monte Carlo denoising under low spp. The core idea of our approach is to reduce the throughput of neural networks by introducing a kernel construction module. Instead of directly predicting per-pixel filtering weights, we treat the neural network as an encoder to output a compact single-channel format of the filtering weights, denoted as importance maps. Then, we feed the importance maps to a hand-crafted decoder to construct the complete kernel map.  In practice, we utilize an unfolding operation and a normalization step to build up the filtering kernel construction module and use the end-to-end training manner to make the network cooperate with this hand-crafted decoder. As a final step, we use the constructed filtering kernels to filter the noisy image, which can be efficiently implemented as screen-space post-processing. 
Our approach benefits from the high parallelism of kernel-based reconstruction and is easy to integrate into existing Monte Carlo denoising modules.
 

\revisionFan{The decoding architecture we designed has the ability to construct specific-sized filtering kernels .}
We further improve the capacity of the decoder with a kernel fusion module. Specifically, the improved decoder constructs filtering kernels from a set of importance maps, then independently applies the kernels to the noisy input and combined the denoised results with learned weights. Practically we construct filtering kernels of different sizes to adapt to various noise frequencies. In addition, we add a temporal accumulation operation before the denoising phase just the same as \cite{10.1145/3269978} does. The increasing effective spp counts of the input image can help stabilize the denoising procedure for consecutive frames and suppress temporal artifacts. 

Comprehensive experiment results show that the proposed approach is good at denoising 1-spp frames at a real-time frame rate. 
\revisionFan{
Our method can preserve kernel prediction methods' denoising quality while roughly halving its denoising time. In addition, compared with the recent neural bilateral grid-based real-time denoiser \cite{.20201133}, our approach benefits from the high parallelism of kernel-based reconstruction and produces better denoising results at equal time.
}
To summarize, our approach makes the following contributions:

\begin{itemize}
\item A novel kernel prediction architecture that uses a compact representation of the kernel map and a scalable decoder for the filtering kernel reconstruction, which significantly reduces neural network throughput \revisionFan{and memory requirement}.
\item \revisionFan{Achieving comparable rendering quality using only roughly half of denoising time compared with the state-of-the-art real-time kernel prediction Monte Carlo denoising method.}
\end{itemize}

\section{Related Work}
\label{sec:related}
Monte Carlo denoising is a long-standing computer graphics research topic in both industrial productions and academic studies, and we refer to the survey from Zwicker et al. \cite{zwicker2015recent} for a complete overview.  In this section, we will mainly discuss deep learning and real-time denoising techniques, which are most related to our method. For a comprehensive study of deep learning-based Monte Carlo denoising techniques, we refer to the recent survey of Huo et al.~\cite{huo2021survey}.

\noindent \textbf{Image-space Denoising.} 
The filtering-based methods are based on using the auxiliary feature buffers to guide the construction of image-space filters \cite{rushmeier1994energy, mccool1999anisotropic, 10.1145/2167076.2167083}. Li et al. \cite{li2012sure} proposed to use the measurements of SURE metric \cite{stein1981estimation} to construct the cross-bilateral filter. Besides, the non-local means filter\cite{buades2005review, Kalantari12} and a joint filtering scheme \cite{rousselle2012adaptive, zimmer2015path} had also been proposed. 

\noindent \textbf{Deep Learning-Based Monte Carlo Denoising.} 
Kalantari et al. \cite{kalantari2015machine} first introduced the supervised learning method for Monte Carlo denoising. They used a multi-layer perceptron (MLP) to estimate the parameters of a fixed-function filter (e.g., cross-bilateral filter). They also computed a rich set of secondary features as inputs to improve the quality. 
Recent state-of-the-art denoising techniques have leveraged the convolutional neural networks (CNN) and noise-free auxiliary feature buffers to reconstruct Monte Carlo renderings \cite{huo2021survey}. Bako et al. \cite{10.1145/3072959.3073708} proposed the kernel prediction method, which utilizes a CNN to predict the weights of per-pixel filtering kernel, and this allows the filtering kernels to be more complex and robust. However, predicting large filtering kernels is time-consuming and memory-exhausting, while using a small-size kernel leads to a quality decrease. Given that filtering with a small-size kernel in down-scaled resolution corresponds to filtering in a large receptive field of the original resolution, Vogels et al. \cite{10.1145/3197517.3201388} proposed a multi-resolution filtering architecture to approximate the behavior of a large kernel. They also decomposed the denoising pipeline with task-specific modules, independently extracting the source-aware and spatio-temporal information. Dahlberg et al. \cite{10.1145/3306307.3328150} deployed this modular hierarchical kernel prediction method to the existing commercial Monte Carlo path tracing engines and showed its practical ability and flexibility in denoising the feature film productions. However, the original kernel prediction method is designed mainly for offline applications with 16-64 spp and runs in seconds. Hasselgren et al. \cite{hasselgren2020neural} and Munkberg et al. \cite{munkberg2020neural} used the hierarchical kernel prediction architecture to denoise the re-sampled Monte Carlo images and the samples-splatted layers, respectively, and they achieved an interactive speed. 
\revisionFan{
Besides, Thomas et al. \cite{10.1145/3414685.3417786} also utilized the hierarchical architecture with a feature extraction network, which is resilient to quantization errors, to explore the feasibility of a heavily quantized network for image reconstruction.
}
Unlike them directly using the kernel prediction architecture, our approach extends it to real-time denoising with 1-spp input by operating on the encoding of the kernel map to reduce neural network inference overhead.

Besides, Xu et al. \cite{xu2019adversarial} proposed to use an adversarial learning approach and emphasize the guidance of feature buffers with a novel conditioned auxiliary feature modulation method. Huo et al.~\cite{huo2020adaptive} denoised incident radiance fields to guide unbiased path tracing. Kettunen et al. \cite{kettunen2019deep} utilized the CNN to replace the screened Poisson solver for gradient-domain path tracing and notably improved the image quality with the help of gradient samples. Gharbi et al. \cite{10.1145/3306346.3322954} proposed a kernel-splatting architecture to estimate the contribution of each raw Monte Carlo sample. This sample splatting method is more natural and robust than the pixel gathering manner in denoising some specific visual effects, but the drawback is that it has a linear complexity of performance and memory with the sample count. Munkberg et al. \cite{munkberg2020neural} extended this method by splatting samples to several layers, thus converting the sample-based computation to layer-based. They take only a fraction of computational cost and memory requirements than the previous per-sample method and preserve a similar quality. Other researches include analyzing path space features to perform manifold contrastive learning to enhance the rendering effect of reflections \cite{cho2021weakly}, filter and reconstruct high-dimensional incident radiation fields for unbiased reconstruction Drawing guide \cite{huo2020adaptive}, etc.

\noindent \textbf{Real-time Denoising.}
Recently, many real-time Monte Carlo reconstruction methods have been proposed. They generally take 1-spp noisy data as input and denoise in a tough time budget. 
\revisionFan{
Mara et al. \cite{mara17towards} used approximations of the bilateral filter,
}
and Schied et al. \cite{schied2017spatiotemporal} used a hierarchical filter expanded with a customized edge-stopping function to filter the temporally accumulated frames progressively. Schied et al. \cite{schied2018gradient} further estimated temporal gradient to guide the temporal accumulation factors. Koskela et al. \cite{10.1145/3269978} applied augmented QR factorization and stochastic regularization to image blocks to do a block-wise feature regression, and their method runs extremely fast under GPU implementation. All of the above methods rely on accumulating projected frames \cite{10.1145/1618452.1618481} to obtain a greater effective spp, which inspires us to incorporate temporal accumulation as our data preprocessing operation.

\revisionFan{Chaitanya}
et al. \cite{10.1145/3072959.3073601} proposed to use a U-Net \cite{ronneberger2015u} convolutional neural network architecture to predict the pixel colors directly. 
\revisionFan{
Hasselgren et al. \cite{hasselgren2020neural} proposed a neural adaptive sampling and denoising architecture with a spatio-temporal joint optimization, where the temporal optimization helps the sample predictor to learn spatio-temporal sampling strategies. As a result, they can achieve real-time rates with optimized network architecture. Compare with their adaptive sampling framework, our method focus on improving kernel-predicting modules' denoising performance and yields faster speed in low-spp (e.g., 1-spp) configurations. 
Besides, Hofmann et al. \cite{hofmann2021interactive} also utilized the neural temporal adaptive sampling architecture with a novel sparse voxel tree data structure to render participating media. 
}
Meng et al. \cite{.20201133} integrated a differentiable bilateral grid to CNN architecture to get a neural bilateral grid denoiser. They gather adjacent pixels to a 3D bilateral space with a learned mapping function and then denoise in this 3D space at real-time frame rates. 
\revisionFan{
\sout{
Their method achieves better quality than previous methods at a comparable performance, which we consider the state-of-the-art real-time Monte Carlo denoising method. 
}
}
However, the bilateral grid-based reconstruction suffers heavy computational overhead compared with our kernel-based reconstruction approach, as demonstrated in Figure~\ref{fig:NBGD_Ours_time}.

\section{\label{sec:problem_statement}Problem Statement}

Our goal is to reconstruct noise-free images from 1-spp path traced images in real-time frame rates, and we achieve this with a supervised learning method. We first generate a set of Monte Carlo path tracing data $\mathcal{S} = ((r_1, \textbf{f}_1, t_1), ..., (r_N, \textbf{f}_N, t_N))$ where $r$ stands for noisy image rendered with low spp, $\textbf{f}$ is the noise-free auxiliary feature (e.g. albedo, normal, depth) obtained in the rendering process, $t$ is the reference image with high spp. We use a denoising function $\mathcal{D}$ with parameters $\Theta$ to reconstruct the denoised image. We 
\revisionFan{notate the}
loss function $\mathcal{L}$ measuring the difference between the denoised image and its reference image, and then minimize the loss function with gradient descent algorithm across the dataset $\mathcal{S}$ with $N$ samples to get the optimal parameters $\Theta_\text{opt}$:
\begin{equation}
\Theta_\text{opt} = \arg~\min_{\Theta} \sum^N_{i=1} \mathcal{L}(\mathcal{D}_\Theta (r_i, \textbf{f}_i), t_i).
\end{equation}
In this paper, the denoising function $\mathcal{D}$ stands for our weight sharing kernel prediction approach, and the trainable parameters $\Theta$ refer to the convolutional neural network weights.

Our approach extends kernel prediction convolutional network (KPCN) \cite{10.1145/3072959.3073708} with a kernel map decoding stage. Directly predicting per-pixel large-size filtering kernel weights spends too much time in network inference for the real-time application. A solution is to 
\revisionFan{prune}
the hierarchy kernel prediction architecture \cite{10.1145/3197517.3201388} for real-time application. However, the solution suffers serious quality decrease. We observe that the total 
\revisionFan{denoising} 
time of the original kernel prediction network is hard to shrink, mostly because of the high throughput of the last convolutional layer. For example, there are $169$ convolutional kernels in the output layer for predicting $13 \times 13$ filtering kernels , which is too heavy for a real-time architecture. Given this, we propose to predict the encoded compact representation of the kernel map and then use an efficient and scalable decoder to reconstruct the complete kernel map. In this manner, we can avoid predicting a heavy $k \times k$ channels kernel map for a filtering window size $k$. Instead, we only need to predict a single-channel kernel map encoding, which dramatically reduces the network throughput. With an efficient and scalable decoder, we can use a lightweight architecture of neural network encoder 
to make our denoising pipeline run at real-time speed while preserving the denoising quality.

Specifically, our neural network $\mathcal{C}$ takes noisy image $r$ and auxiliary feature $\textbf{f}$ as input, and then estimates a single scalar for each pixel $p$, which we denote as importance term and the whole single-channel image is denoted as importance map:
\begin{equation}
\label{function1}
I(p) = \mathcal{C}_{\Theta}(r, \textbf{f})(p).
\end{equation}
This 2D single-channel importance map is the predicted encoding of the kernel map. We then use it to construct the filtering kernel with a hand-crafted decoder. Letting $\Omega_p$ denote the $k \times k$ neighborhood centered around pixel $p$, which is also the filtering window for $p$, we compute the normalized filtering kernel weight $w_p(q)$, where $q \in \Omega(p)$, in a splatting manner:
\begin{equation}
w_p(q) = \frac{\text{exp}(I(q))}{\sum_{q^{'} \in \Omega_{p}} \text{exp}(I(q^{'}))}.
\end{equation}
The softmax normalization used here is a common practice \cite{10.1145/3072959.3073708, hasselgren2020neural} to enforce that $0 \leq w_p(q) \leq 1$ and it contributes to a more steady training process. Then, we apply the same weights to each RGB color channel and compute the denoised pixel color as
\begin{equation}
R(p) = \sum_{q \in \Omega_p} w_p(q) r(q).
\end{equation}

The filtering kernel size $k$ is pre-defined along with the other hyperparameters (e.g., CNN layer count, convolutional kernel size), and this means the importance term we estimated is a relative value for a given filtering window size. 
Inspired by Vogels et al. \cite{10.1145/3197517.3201388} who used same sized filtering kernels in different resolutions to improve the denoising quality, we design a kernel fusion module by filtering with different sized kernels in a same resolution, which has also improved our decoder's scalability. Precisely, we predict $M$ importance maps
\Skip{
We further design a kernel fusion \ren{kernel embedding?} 
\cite{10.1145/3197517.3201388, hasselgren2020neural, munkberg2020neural} all used 'same size, multi resolution', which is different from our 'multi sizes, same resolution' setting, is it enough to cite one of them? \huo{one paper is enough.}
}
to construct filtering kernels with different filtering window sizes $k_i$. Then we independently filter the noisy input and average the results $R^{k_i}$ with the learned weights, $\alpha_i(p)$, as: 
\begin{equation}
\label{equa:kernel_fusion}
\hat{R}(p) = \sum_{i=1}^{M} \alpha_i(p) R^{k_i}(p),
\end{equation}
where $0 \leq \alpha_i(p) \leq 1$ and $\sum_{i=1}^M \alpha_i = 1$. 
\Skip{
\huo{talk about the motivation before giving the equation.} \fan{should we give the motivation of the average method we used (weighted average)? Actually \cite{hasselgren2020neural} does use a special average method, which is an ordered alpha-blending, to combine their independently filtered layers. But in my opinion it is mostly because their layers have no special semantic meanings with each other, so this ordered alpha-blending can manually give the layers some semantic meanings like different layers correspond different depths. However, our independent filtering kernels already have semantic meanings, i.e. the kernel size, which is considered to correspond to different frequency, so I think we do not need special average method, and a simple weighted average is enough. Is this analysis reasonable and should we add this to the article?} \huo{The difference of weighting scheme seems to be technical details that too subtle for Sec.3. Let's skip it for now.} 
}
This kernel fusion module helps our denoiser distinguish the frequency of noise region and essentially improves the scalability of our decoder architecture.

\section{Weight Sharing Kernel Prediction Architecture}

\begin{figure*}[htb]
  \centering
  \includegraphics[width=1.0\linewidth]{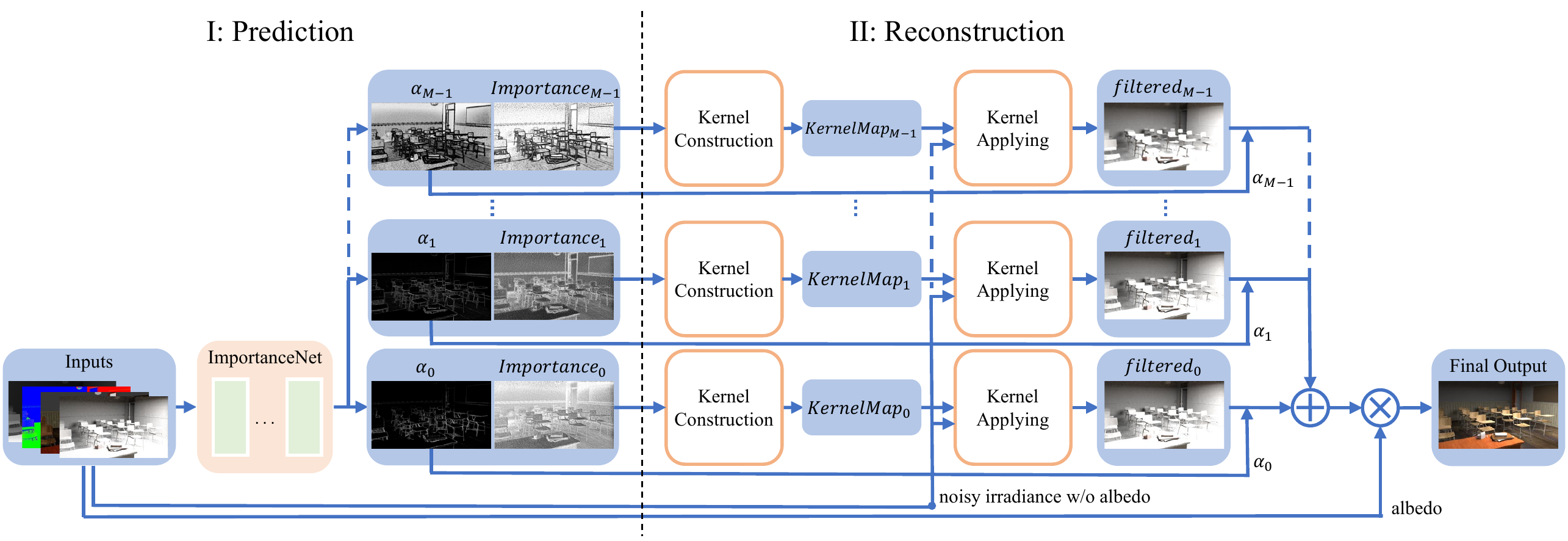}
  \caption{\label{fig:pipeline}
            An overview of our kernel prediction denoising framework. The input data are preprocessed and sent to the ImportanceNet, and we construct the filtering kernels with varying sizes from the evaluated importance maps. Then we fuse the independently filtered images with a weighted average and multiply back the albedo to obtain the final denoised image.}
\end{figure*}

In this section, we describe the overall architecture of our weight sharing kernel prediction neural denoiser. As illustrated in Figure~\ref{fig:pipeline}, our architecture consists of the prediction phase and the reconstruction phase. At the prediction phase (Section~\ref{sec:prediction_phase}), the input data are sent to an \textit{ImportanceNet} to predict the importance map, i.e., 
\revisionFan{a} 
compact encoding of the kernel map. At the reconstruction phase (Section~\ref{sec:reconstruction_phase}), we construct the filtering kernels from the importance map to filter the noisy input. Besides, we design a kernel fusion architecture that constructs filtering kernels with different window sizes to filter the input independently, and then fuses the results with a weighted average.

\subsection{\label{sec:prediction_phase}Prediction Phase}


\begin{figure}[htb]
  \centering
  \includegraphics[width=1.0\linewidth]{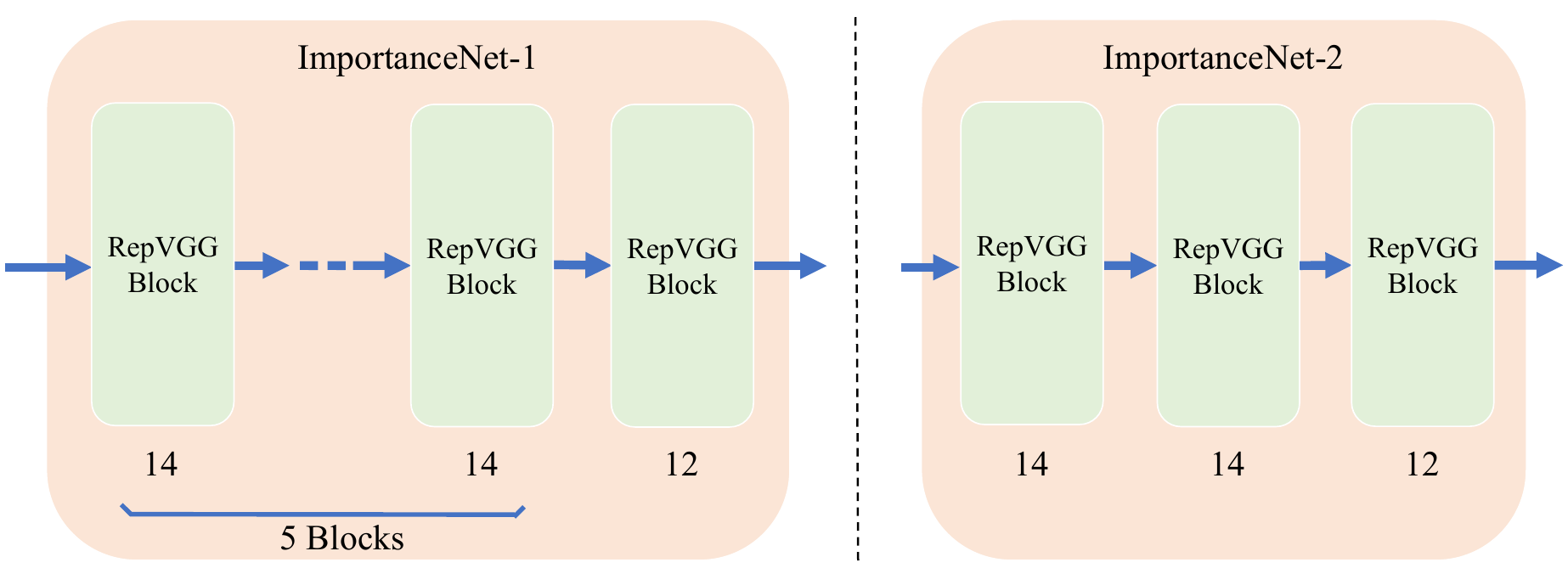}
  \caption{\label{fig:ImportanceNet}
           Architectures of ImportanceNet for 1-spp input. One has six RepVGG Blocks (left) and another employed three (right). The output channels of each layer is shown below each block. The structure of RepVGG is in the supplementary document.
           }
\end{figure}

At the prediction phase, we use a CNN, denoted as \textit{ImportanceNet}, to predict the encoding of the kernel map. Our \textit{ImportanceNet} takes the 1-spp path traced image and its auxiliary feature buffers as inputs. It outputs a single-channel importance map, which is used for constructing the filtering kernel map at the reconstruction phase. We build our \textit{ImportanceNet} with a variant of \textit{RepVGG Block} structure \cite{2021arXiv210103697D}. Experimentally we use two network architectures to denoise 1-spp input: the first one employs six \textit{RepVGG Blocks} for a better quality (Figure~\ref{fig:ImportanceNet}, left) and another employs three for a faster execution speed (Figure~\ref{fig:ImportanceNet}, right). They all run at real-time speed. 


The \textit{RepVGG Block} structure \cite{2021arXiv210103697D} can improve the capacity of a fully convolutional neural network during training and will not introduce additional time cost to network inference, and this is implemented with a parameter conversion. One \textit{RepVGG Block} has multiple branches at training-time: $1 \times 1$, $3 \times 3$, $5 \times 5$ convolution layer (Conv) branch, and identity branch. These branches are combined with an element-wise addition and sent to a ReLU layer. After training, the multi-branches structure is equivalently transformed to a single $5 \times 5$ Conv layer for inference. Please see the supplementary document for the details of this structure. The converted network architecture is an efficient single-branch fully convolution network and does the same computation before conversion. The over-parameterization property of \textit{RepVGG Block} is practically beneficial to training, please see Section~\ref{sec:ablation_study} for a detail comparison.

Besides, temporally accumulating consecutive 1-spp frames can effectively improve the temporal stability and increase the effective spp of each frame. We employ a temporal accumulation preprocessing step before sending the noisy inputs to the denoising pipeline just like \cite{schied2017spatiotemporal, 10.1145/3269978, .20201133}. We first reproject the previous frame to the current frame with the motion vector and then judge their geometry consistency by world position and shading normal feature buffers. Current frame pixels that passed the consistency test are blended with their corresponding pixels in the previous frame, while the failed pixels remain original 1 spp.

\subsection{\label{sec:reconstruction_phase}Reconstruction Phase}

\begin{figure}[htb]
  \centering
  \includegraphics[width=1.0\linewidth]{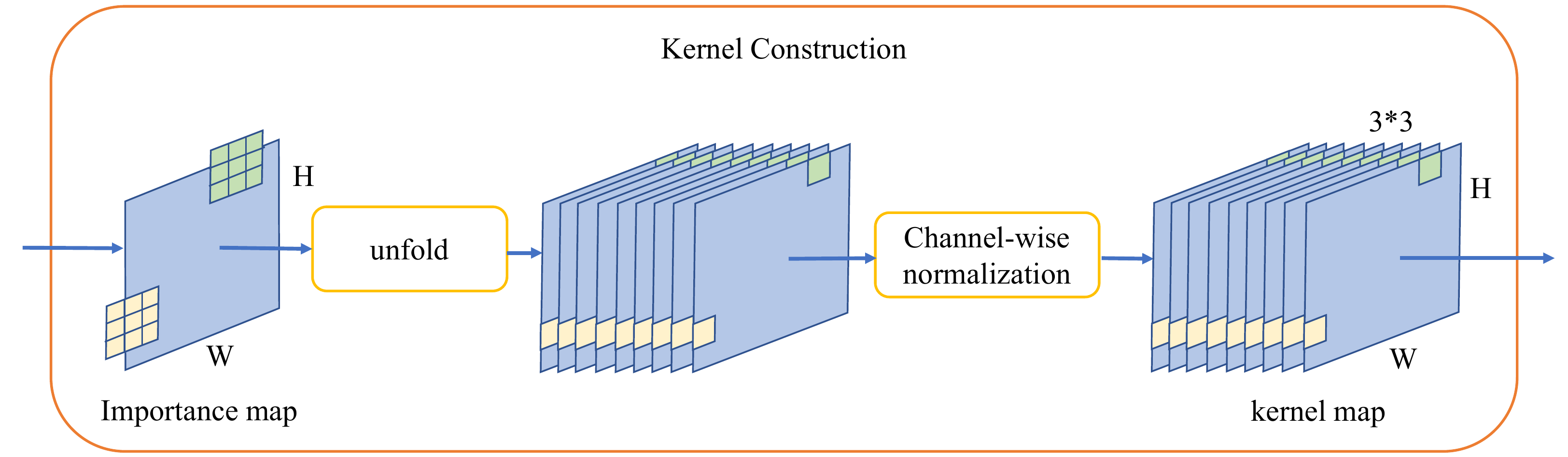}
  \caption{\label{fig:kernel_construction}
           Framework of kernel construction module. We construct a kernel map with $3 \times 3$ filtering window size for illustration. For the input single-channel importance map, we use a sliding window of size $3$ to unfold it to 9 channels in a splatting manner, then we apply a channel-wise normalization to obtain the kernel map. The yellow and green color cells represent data flow of two example positions.
          }
\end{figure}

\noindent \textbf{Kernel Construction.}
The original kernel prediction method \cite{10.1145/3072959.3073708} directly predicts the kernel map with a resolution $H \times W \times (k * k)$ for the per-pixel filtering kernels of window size $k \times k$. While achieving high quality, this scheme leads to a heavy inference overhead when predicting a large-size kernel, limiting its application in real-time application. 
In order to address this limitation, our method predicts the encoding of the kernel map with a purpose $H \times W \times 1$, namely the importance map. Then we manually construct the $H \times W \times (k*k)$ kernel map from the encoding. The kernel construction process is illustrated in Figure~\ref{fig:kernel_construction}. For each pixel, we leverage the neighborhood scalar values in the importance map to construct its filtering kernel weights. Specifically, we first unfold the importance map with a sliding window with window size $k$ to obtain the unnormalized kernel map with resolution $H \times W \times (k*k)$. Then we normalize it with a softmax function along the channel axis. The softmax normalization helps our constructed filtering kernels have conserving energy, and it ensures well-behaved gradients for the network training \cite{10.1145/3072959.3073708, hasselgren2020neural}. The constructed kernel map stores filtering kernel weights for each pixel, which are used to filter the noisy input. Note that there 
\revisionFan{are}
no learnable parameters in our kernel construction process.

As mentioned above (Section~\ref{sec:problem_statement}), we extend the kernel prediction method to a real-time architecture by enabling a lightweight network design with a scalable kernel construction module (Figure~\ref{fig:pipeline} II) 
For the kernel construction part, we observe negligible time cost compared with the total prediction module (Figure~\ref{fig:pipeline} I),
attributing to the efficient kernel construction scheme. Note that our reconstruction module has a constant inference time with respect to the size of the kernel map. Hence, our method runs faster in constructing large-sized kernel maps compared with the original kernel prediction method.

\noindent \textbf{Kernel Fusion.}
\revisionFan{While filtering kernels are widely varying, encoding very complex filtering information into only one single-channel importance map has limitations because the filtering weights are highly correlated for nearby pixels. For example, it might not be able to capture signal changes in different directions around object edges, while this could be achieved with KPCN \cite{10.1145/3072959.3073708} by independently estimating a separate kernel at each pixel. Our alternative solution is to reconstruct multiple channels of importance maps at each pixel to decompose directional filtering information then fuse the denoised to achieve the final results. 
}
Hence, we further design a kernel fusion architecture to improve the denoising ability of our method. Specifically, given the scalability of our \textit{ImportanceNet} and the reconstruction module, we can decompose the filtering information to multiple importance maps to encode more information. As shown in Figure~\ref{fig:pipeline}, instead of predicting only one importance map and constructing one \textbf{fusing kernel}, our \textit{ImportanceNet} predict multiple separate importance maps, and then we construct a set of kernel maps with different filtering sizes $k_i$ and $i \in \{0, ..., M-1\}$. We independently filter the noisy image with filtering kernels of different sizes and then fuse the results with a weighted average to form the final filtered result
\revisionFan{
as in Equation~\ref{equa:kernel_fusion}, where the per-pixel average weights $\alpha_i$ are also predicted by our \textit{ImportanceNet} in the last convolutional layer and normalized by a softmax activation function. In our implementation, we use a sequence of even-sized ($\{3, 5, 7, ...\}$) filtering kernels, which is a common selection of kernel size like that in \cite{2021arXiv210103697D}. Please see Section~\ref{sec:ablation_study} for the evaluation of our kernel fusion module. }

\revisionFan{Furthermore, with}
the kernel fusion architecture, filtering kernels of different sizes respond to filtering on different frequencies. Our averaging weight map is trained to blend the filtered results on each pixel independently. It gives higher weight to large-sized kernel filtered results in low-frequency regions and 
\revisionFan{vice}
versa. While fusing more filtering kernels could improve the denoising quality, it costs more time.  We need to choose an appropriate fusing kernel count to have a good trade-off between quality and performance. Ablation study for the number of kernels in Section~\ref{sec:ablation_study} is conducted for the appropriate number we chose in our implementation.

As shown in Figure~\ref{fig:pipeline}, we filter the noisy input irradiance without albedo, and at the last step, we multiply back the albedo to get the final denoised radiance image. 
\revisionFan{
The albedo demodulation and re-modulation is a common practice \cite{zimmer2015path, 10.1145/3072959.3073708, 10.1145/3269978, .20201133} because the effective irradiance buffer is smoother and has simpler noise patterns. 
}
Thus denoising with the irradiance can avoid over-blurring the detail of the texture.

\section{\label{experimental}Experimental Setup}

\subsection{Datasets}

We adopt an existing dataset from the work of Koskela et al. \cite{10.1145/3269978} called the BMFR dataset. This dataset comprises 1-spp path traced noisy images, 4096-spp reference images, and the related feature buffers (albedo, shading normal, word position, and camera-space depth) from six scenes 
with rendering effects like glossy reflections and soft shadows. Each scene has 60 consecutive frames with smooth camera movement. The similar setup is also adapted by Meng et al. \cite{.20201133}. 

\begin{figure}[htb]
  \centering
  \includegraphics[width=1.0\linewidth]{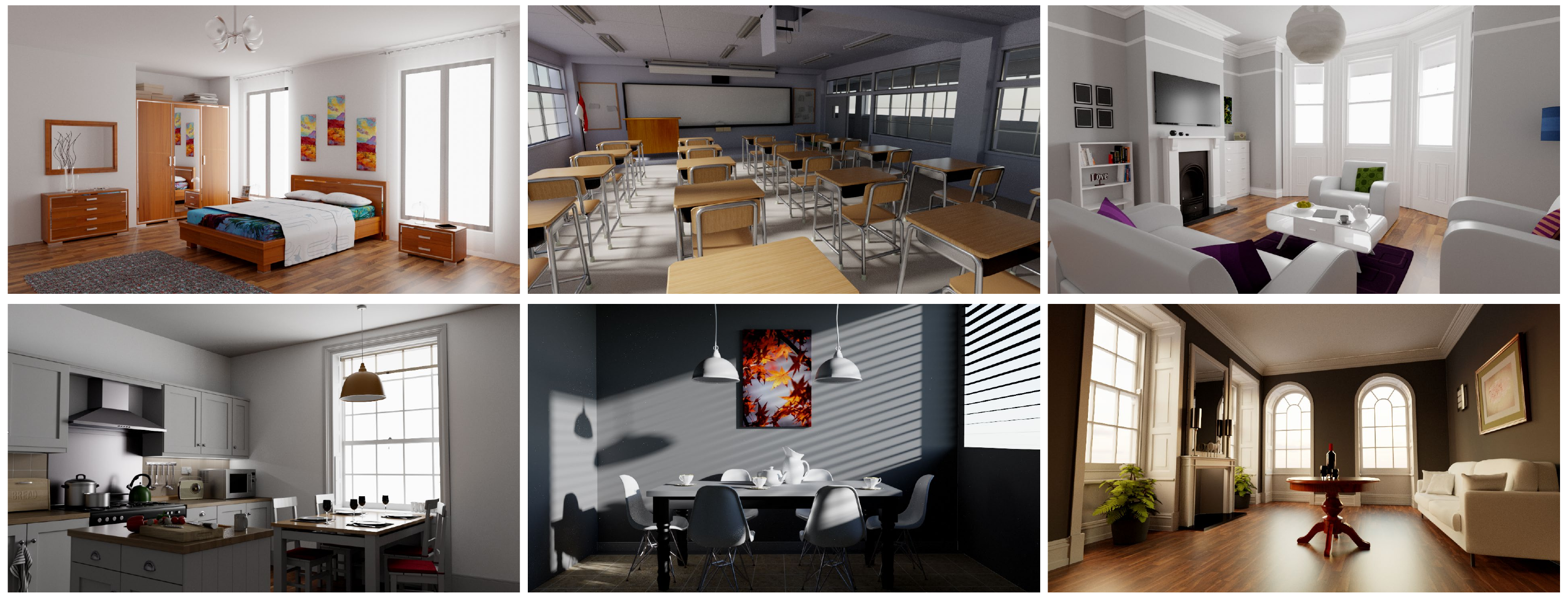}
  \caption{\label{fig:tungsten_overview}
            Example scenes of Tungsten dataset. The reference images are rendered with 4096 spp.
          }
\end{figure}

To evaluate our generalization ability to high spp input, we also use another high-quality Tungsten dataset made public by Meng et al. \cite{.20201133}. This dataset is rendered with eight publicly available Tungsten scenes \cite{resources16} covering complex geometry information and lighting conditions. See Figure~\ref{fig:tungsten_overview} for some example scenes. Each scene contains a consecutive sequence of 100 frames, and the noisy image is rendered with 64 spp. 

Our inputs to the \textit{ImportanceNet} are comprised of low dynamic range (LDR)  3-channel 
\revisionFan{color}
modulated with 3-channel albedo, 3-channel normal, and 1-channel depth (10-channel in total). We linearly scale normal buffer and depth buffer to range $[0, 1]$.
 We apply tone mapping for the original high dynamic range (HDR) input radiance (gamma correction for the BMFR dataset and Filmic tone mapping for the Tungsten dataset) before it is sent to the \textit{ImportanceNet}. In the kernel applying step, we filter the HDR irradiance without albedo to preserve the original value distribution.

For both datasets, we hold out frames of one scene as test set and use the other scenes as the training data. During training, we uniformly sample $80$ image patches with resolution $128 \times 128$ from each $1280 \times 720$ frame of training data to form a training dataset and sample another $20$ patches each frame for the validation dataset.

\subsection{\label{lab:training} Training}

For the BMFR dataset, 
\revisionFan{
we use the \textit{symmetric mean absolute percentage error} (SMAPE) \cite{10.1145/3197517.3201388} as suggested by \cite{10.1145/3197517.3201388, munkberg2020neural}. Given denoised image $R$ and reference $t$ in HDR domain, SMAPE computes:
}

\begin{equation}
SMAPE(R, t) = \frac{1}{3N} \sum_{p \in N} \sum_{c \in C} \frac{|R_{p, c} - t_{p, c}|}{|R_{p, c}| + |t_{p, c}| + \epsilon},
\end{equation}
where N is the number of pixels in the image, $C$ denotes color channels, and $\epsilon$ is a small number $0.01$. 


For the BMFR dataset 
, we use \textit{RepVGG Block} to build our \textit{ImportanceNet} as described in Section~\ref{sec:prediction_phase} for real-time speed. For the Tungsten dataset, which is not used for real-time application, we use a more complex network with multi-resolution architecture \cite{10.1145/3197517.3201388, .20201133} (the network architecture details can be found in the supplementary document). 
This network has a U-net structure \cite{ronneberger2015u}, and we construct kernels and filter the input in 3 resolutions. For two adjacent resolutions, we denote the coarse-resolution image, fine-resolution image, and the predicted pixel-wise blending weight with $\textbf{i}^c$, $\textbf{i}^f$, and $\alpha$, respectively. $\textbf{D}$ and $\textbf{U}$ are $2 \times 2$-downsampling and nearest-neighbor upsampling operators, respectively. Then the denoised results of different resolutions are combined progressively from the coarsest resolution as:
\begin{equation}
\textbf{o}_p = \textbf{i}^f_p - \alpha_p \left[ \textbf{U} \textbf{D} \textbf{i}^f \right]_p + \alpha_p \left[ \textbf{U} \textbf{i}^c \right]_p.
\end{equation}

\subsection{Implementation}
Our denoiser is implemented in PyTorch \cite{paszke2017automatic}. We implemented filtering kernel construction and fusion modules in CUDA as extended operators of PyTorch. 
\revisionFan{ We use one single function to realize kernel construction, filtering, and fusing in a streaming manner, which avoid explicitly constructing very large kernel maps stored in global memory as the basic KP method does. In this manner, the memory footprint when denoising one 720p-frame is reduced from 3.95G to 1.25G compared with KP.}
The weights of our neural network are initialized with the default uniform distribution. We use a batch size of $64$ and train the network with Adam \cite{kingma2014adam} optimizer with a learning rate of $0.001$. For the BMFR dataset, we construct six filtering kernels for kernel fusion module, which means that our \textit{ImportanceNet} predicts 6 importance maps and six corresponding averaging weight maps. We use base kernel size $k_b = 3$ and kernel size incremental step $k_s = 2$, so our minimum kernel size is $3$ and maximum kernel size is $13$. For the Tungsten dataset, we fuse two filtering kernels with sizes 3 and 5 for each resolution. For each test case, we hold out one scene as test set and train the network with remaining scenes for 500 epochs, which takes 12 hours on an NVIDIA RTX 2080 Ti GPU. 

\section{Results}
\subsection{Evaluation Metrics and Compared Algorithms}

We use Peak Signal-to-Noise Ratio (PSNR) and Structural SIMilarity (SSIM) \cite{wang2003multiscale} as metrics to evaluate our results. We compare our method to a range of existing real-time denoisers: Neural Bilateral Grid Denoiser (NBGD) \cite{.20201133}, BMFR approach \cite{10.1145/3269978}, OptiX Neural Network Denoiser (ONND)~\cite{10.1145/3072959.3073601}, Spatio-temporal Variance-Guided Filtering method (SVGF) \cite{schied2017spatiotemporal}, and two variants of Kernel Prediction Convolutional Network \cite{10.1145/3072959.3073708} (KPCN) with a small network size (KP) or a Multi-Resolution architecture (MR-KP)~\cite{10.1145/3197517.3201388} 
modified by Meng et al. \cite{.20201133}. We also include a comparison with the traditional offline filter-based denoiser NFOR \cite{bitterli2016nonlinearly}. We use the implementations provided by their authors except for ONND and KP.

We use two KPCN variants for a comprehensive evaluation. The first variant KP has the same lightweight network architecture as ours (six \textit{RepVGG Blocks}) and predicts filtering kernels with size $13 \times 13$. The second variant MR-KP adapted by Meng et al. \cite{.20201133} has a similar multi-resolution network architecture as described in Section~\ref{lab:training} and predicts filtering kernels with size $5 \times 5$ in three resolutions. We trained the variants using the same loss function and dataset as ours, and both of them run at interactive frame rates.


Following the setup of the original paper, we evaluate the NBGD \cite{.20201133} with two convolutional layers for the BMFR dataset and seven convolutional layers with dense connections for the Tungsten dataset, respectively.
Both the network architectures have the same multi-scale framework with three bilateral grids. We keep it the same as the original paper when doing the experiments. 

ONND \cite{10.1145/3072959.3073601} is provided as a black box module in OptiX 5.1, and it performs albedo demodulation and re-modulation in the module itself. Because the neural network model of ONND is trained with toned mapped 1-spp input data without temporal accumulation, we send input data without temporal accumulation to avoid introducing bias.

For NBGD, BMFR, SVGF, MR-KP, and KP, we use the same temporal accumulation and albedo demodulation and re-modulation preprocessing steps as ours. Because we do not observe improvement on NBGD with the SMAPE loss, we keep the original loss function of NBGD, i.e. $L_1$,  unchanged to achieve its best for a fair comparison.

\subsection{Quality Comparisons}

We present some of the visual results and the average PSNR and SSIM in this section. Please see our supplementary document for more comparisons using other error metrics.


\begin{figure*}[htb]
  \centering
\includegraphics[width=1.0\linewidth]{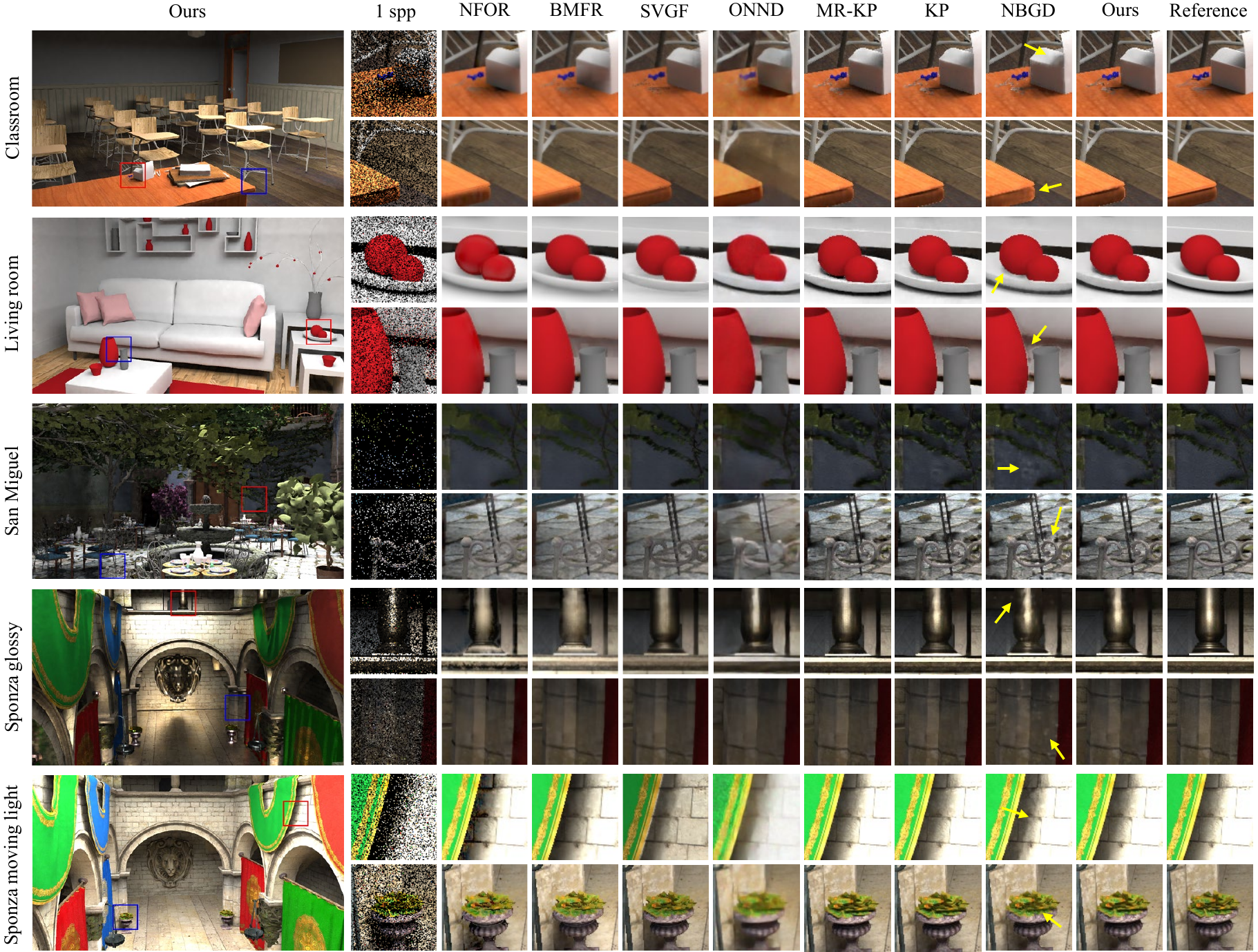}
  \caption{\label{fig:BMFR_results}
           Visual comparisons of denoising quality on the 1-spp BMFR test data. Each row represents an independent experiment where the displayed scene is held up for test and other scenes are used for training. Closeups highlight the visual differences between our method and other compared methods. 
           }
\end{figure*}

\noindent \textbf{BMFR dataset.}
We use a lightweight 6-layer convolutional neural network and six fused filtering kernels to denoise the 1-spp BMFR dataset in real-time speed. The example denoised images in test scenes are shown in Figure~\ref{fig:BMFR_results}. 
\revisionFan{
Overall, kernel prediction-based methods (KP, MR-KP) produced visually pleasing results, and our method achieved a similar visual and numerical quality on par with them.
}
Compared with \revisionFan{one of } the state-of-the-art real-time neural \revisionFan{denoisers} NBGD, our method constructed more realistic and clear shadows using less denoising time (Figure~\ref{fig:BMFR_results}, crops row 1, 3, 9, 10). The grid splatting method of NBGD had difficulty suppressing outlier pixels and left some firefly artifacts for glossy material, while our method reconstructed more smooth glossy reflection results (Figure~\ref{fig:BMFR_results}, crops row 7, 8). Besides, our method performed better at preserving object edges and handling occlusions with fewer artifacts (Figure~\ref{fig:BMFR_results}, crops row 2, 4, 5, 6).

\begin{table*}
  \centering
  \caption{
  Average PSNR comparison (higher is better) on 1-spp BMFR test data. 
  We also added the metric averaged with all test scenes at the last row and noted that we use hold-out dataset partition, this averaged metric is just for a rough analysis.
  Ours 6-layer represents the 6-layer convolutional neural network architecture and Ours 3-layer represents the 3-layer convolutional neural network architecture.}
  \label{tab:BMFR_PSNR}
  \begin{tabular}{cccccccccc}
   \toprule
    \multirow{2}{*}{Scene}&
    \multicolumn{9}{c}{PSNR}\\
    \cmidrule(lr){2-10}
                        &  NFOR  &  BMFR  &  ONND  &  SVGF  & MR-KP  &   KP   &  NBGD  & Ours(6-layer) & Ours(3-layer) \\
    \midrule
    Classroom           & 29.872 & 28.965 & 27.312 & 25.034 & 33.030 & \textbf{33.047} & 31.534 &    32.827     &    32.220     \\
    Living room         & 31.304 & 30.025 & 25.586 & 27.239 & 33.832 & \textbf{34.090} & 32.506 &    34.063     &    32.913     \\
    San Miguel          & 21.811 & 20.969 & 20.172 & 18.736 & 24.047 & 24.215 & 23.807 &    \textbf{24.269}     &    23.860     \\
    Sponza              & 30.377 & 31.111 & 24.698 & 24.401 & \textbf{34.730} & 34.595 & 33.412   &    34.600     &    33.796     \\
    Sponza (glossy)     & 25.974 & 25.005 & 23.460 & 20.917 & \textbf{30.923} & 30.719 & 29.678 &    30.805     &    29.981     \\
    Sponza (mov. light) & 21.999 & 17.377 & 22.291 & 17.260 & 25.372 & 25.324 & 24.866 &    \textbf{25.374}     &    25.050     \\
    Average             & 26.889 & 25.575 & 23.920 & 22.264 & 30.322 & \textbf{30.332} & 29.474 &    30.323     &    29.637     \\
    
  \bottomrule
\end{tabular}
\end{table*}

\begin{table*}
  \centering
  \caption{
  Average SSIM comparison (higher is better) on 1-spp BMFR test data. 
  We also added the metric averaged with all test scenes at the last row and noted that we use hold-out dataset partition, so this averaged metric is just for a rough analysis.
  Ours 6-layer represents the 6-layer convolutional neural network architecture and Ours 3-layer represents the 3-layer convolutional neural network architecture.}
  \label{tab:BMFR_SSIM}
  \begin{tabular}{cccccccccc}
   \toprule
    \multirow{2}{*}{Scene}&
    \multicolumn{9}{c}{SSIM}\\
    \cmidrule(lr){2-10}
                        &  NFOR  &  BMFR  &  ONND  &  SVGF  & MR-KP  &  KP    &  NBGD  & Ours(6-layer) & Ours(3-layer) \\
    \midrule
    Classroom           & 0.956  & 0.955  & 0.924  & 0.952  & \textbf{0.978}  & \textbf{0.978}  & 0.967  &    0.977      &    0.973      \\
    Living room         & 0.967  & 0.965  & 0.953  & 0.950  & 0.977  & 0.978  & 0.971  &    \textbf{0.979}      &    0.973      \\
    San Miguel          & 0.799  & 0.789  & 0.744  & 0.790  & \textbf{0.851}  & \textbf{0.851}  & 0.833  &    0.849      &    0.841      \\
    Sponza              & 0.939  & 0.948  & 0.852  & 0.927  & 0.982  & 0.982  & 0.975  &    \textbf{0.983}      &    0.979      \\
    Sponza (glossy)     & 0.901  & 0.907  & 0.867  & 0.913  & \textbf{0.962}  & 0.960  & 0.946  &    0.961      &    0.955      \\
    Sponza (mov. light) & 0.895  & 0.858  & 0.811  & 0.876  & \textbf{0.961}  & 0.958  & 0.947  &    0.958      &    0.953      \\
    Average             & 0.910  & 0.904  & 0.858  & 0.901  & \textbf{0.952}  & 0.951  & 0.940  & 0.951      &    0.946      \\
    
  \bottomrule
\end{tabular}
\end{table*}
\begin{table}
  \centering
  \caption{Run-time performance of each denoising approach at a resolution of $1280 \times 720$. Ours 3-layer, Ours 6-layer, and NBGD 2-layer are designed for 1~spp denoising, while Ours MR and NBGD 7-layer are designed for 64-spp denoising. For neural denoising methods, we decompose the total denoising time into prediction time and reconstruction time, and they are shown in the bracket, respectively. We ran 300 iterations for each method and reported the average timing for one single frame to decrease random fluctuations.
  }
  \label{tab:time}
  \begin{tabular}{ccc}
    \toprule
    Method       & Timing(ms) & Device                \\
    \midrule
    Ours 3-layer & ~~6.58 (~~5.47 / 1.11)       & Nvidia RTX 2080 Ti     \\
    Ours 6-layer & 12.89 (11.75 / 1.14)     & Nvidia RTX 2080 Ti     \\
    Ours MR      & 22.70 (21.85 / 0.85)     & Nvidia RTX 2080 Ti     \\
    NBGD 2-layer & 13.97 (~~7.84 / 6.13)      & Nvidia RTX 2080 Ti     \\
    NBGD 7-layer & 50.28 (44.16 / 6.12)     & Nvidia RTX 2080 Ti     \\
    KP           & 27.86 (26.23 / 1.22)     & Nvidia RTX 2080 Ti     \\
    MR-KP        & 24.73 (23.73 / 1.00)     & Nvidia RTX 2080 Ti     \\
    BMFR         & 1.60                   & Nvidia RTX 2080        \\
    SVGF         & 4.40                   & Nvidia Titan X         \\
    ONND         & 55.00                  & Nvidia Titan X         \\
    NFOR         & 370.00                 & Intel Core i7-8700 CPU \\
  \bottomrule
\end{tabular}
\end{table}


We report the numerical results in Table~\ref{tab:BMFR_PSNR} and Table~\ref{tab:BMFR_SSIM}. The corresponding average execution time is reported in Table~\ref{tab:time}. 
\revisionFan{
Typically, kernel prediction-based methods (KP, MR-KP) have better quantitative quality than other denoising approaches, albeit some of them are faster. Our method achieves a comparable result with the kernel prediction-based methods and runs at a roughly 2-times faster speed. 
}
\revisionFan{
Compared with the recent real-time denoiser NBGD, 
}
our method has lower quantitative errors while runs faster than NBGD (12.9 ms vs. 14.0 ms). Besides, given the scalability of our architecture, we adopted a more efficient \textit{ImportanceNet} with a 3-layer convolutional neural network and six fused filtering kernels  (Figure~\ref{fig:ImportanceNet}, right) for better run-time performance. As shown in Table~\ref{tab:BMFR_PSNR}, \ref{tab:BMFR_SSIM}, and~\ref{tab:time}, the denoising quality of our optimized architecture is on par with NBGD, but ours only needs 6.6 ms to denoise one 720p frame, which is roughly a half of NBGD's 14.0 ms.
\begin{figure}[htb]
  \centering
  \includegraphics[width=1.0\linewidth]{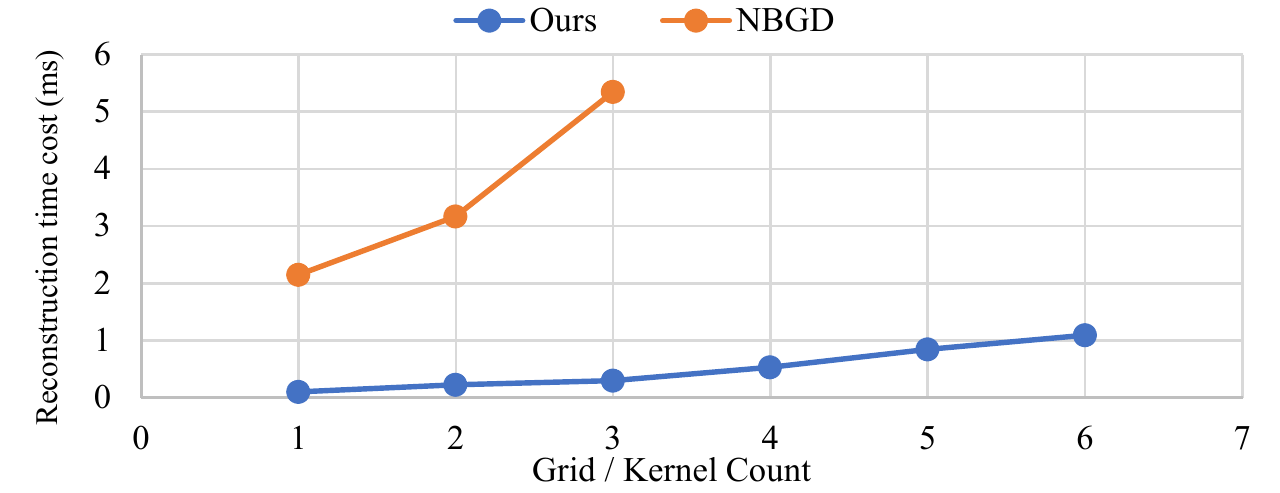}
  \caption{\label{fig:NBGD_Ours_time}
           The reconstruction timing change over guide channel size. For NBGD, the x-axis represents the number of bilateral grid used, and the grid resolutions vary from $320 \times 180 \times 64$ to $80 \times 45 \times 16$ with scale ratio $2$. For our method, the x-axis represents the number of filtering kernel fused, and the kernel sizes vary from $3 \times 3$ to $13 \times 13$ with incremental step $2$. 
           }
\end{figure}

As discussed in Section~\ref{sec:related}, our kernel-based reconstruction scheme is more efficient compared with that of NBGD's. Here we perform an ablated study to compared the time cost of the reconstruction parts of NBGD and ours: the grid creating and slicing of NBGD, and the kernel construction and filtering of our method, precisely. 

As shown in Figure~\ref{fig:NBGD_Ours_time}, the grid-based operations are more time-consuming than our kernel-based operations. For the single resolution architecture and similar reception field comparison, NBGD with a window of size $8$ takes $2.15$ ms while our method of size $9$ takes $0.31$ ms. For the complete architecture comparison, NBGD with $3$ grids takes $5.35$ ms, while our method with $6$ kernels fused takes $1.10$ ms. In addition, the reconstruction time cost of NBGD grows significantly faster over guide channel size (grid count or kernel count) than ours, demonstrating that our approach has better performance for scalable network outputs. Compared with NBGD, the efficiency of our reconstruction part allows our denoising pipeline to use a deeper network under the same time budget.

\begin{figure*}[htb]
  \centering
  \includegraphics[width=1.0\linewidth]{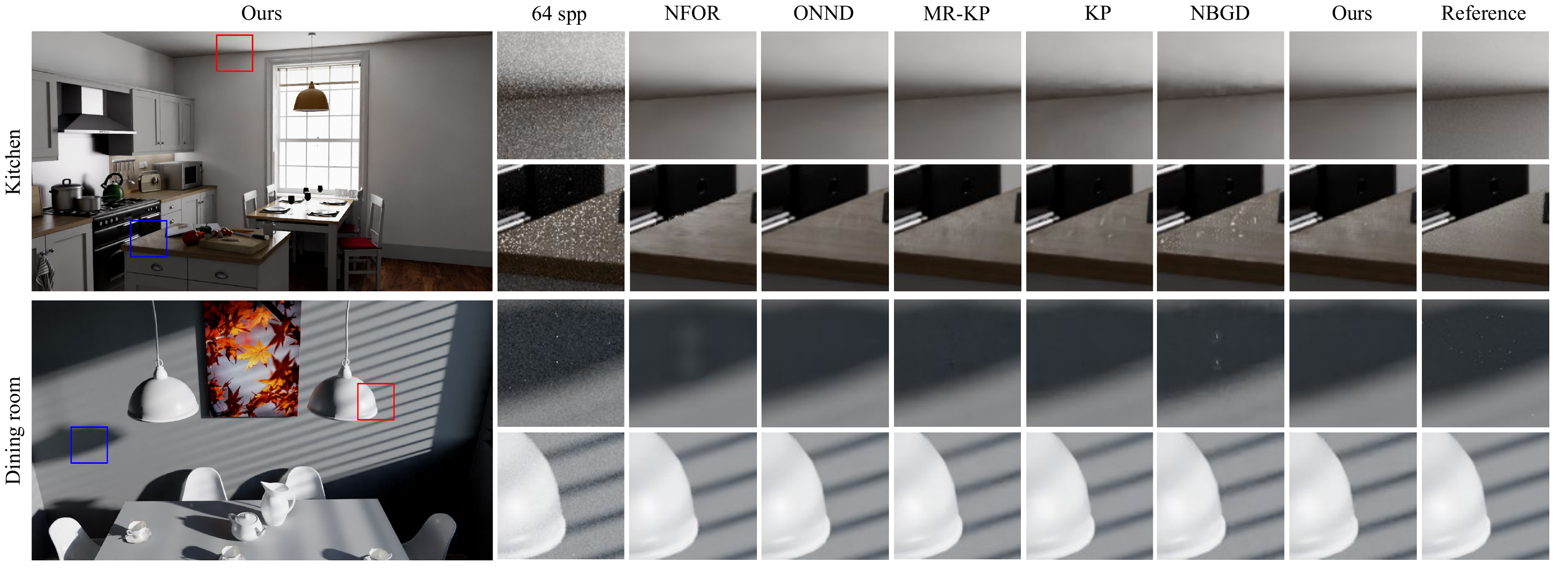}
  \caption{\label{fig:Tungsten_results}
           Visual comparisons of denoising quality on the 64-spp Tungsten test scenes of \textit{Kitchen} and \textit{Dining room}. We use the multi-resolution neural network architecture with two fused kernels in each resolution. 
           }
\end{figure*}
\input{Tungsten_metrics}

\noindent \textbf{Tungsten dataset.}
To denoise the 64-spp Tungsten dataset, we use a multi-resolution neural network architecture with two fused filtering kernels  at each resolution as described in Section~\ref{lab:training}. We show the visual comparisons of denoised results in Figure~\ref{fig:Tungsten_results} and the average error metrics in Table~\ref{tab:Tungsten_metrics}. Note that for the Tungsten dataset, which is generated for offline applications, we do not use the temporal accumulation step. Overall, our method has the denoised results on par with the compared methods in both visual quality and quantitative metrics.


Due to the complex configuration of Tungsten scenes and the limited sampling rates, in both 64-spp noisy input and the 4096-spp reference image, there exist outlier pixels whose values are significantly larger than their neighborhoods. Meng et al. \cite{.20201133} suggested an outlier removal preprocessing step to the noisy input. However, this operation would change the energy distribution of the original path traced image and have a negative influence on error metrics, so we did not include it in our comparison experiments. The blue insets of the \textit{Dining room} (row 2 of Figure~\ref{fig:Tungsten_results}) show that NBGD and NFOR could not effectively remove the outliers while ours, MR-KP, and ONND reduced most outliers to an unnoticeable level. Besides, as shown in the insets of \textit{Kitchen} scene (row 1 of Figure~\ref{fig:Tungsten_results}), our approach excelled at reconstructing high-frequency glossy reflection and smooth shadows, while MR-KP and NBGD still left some residual noise.  


We report the average denoising time of our multi-resolution 2-kernel architecture in the third data row of Table~\ref{tab:time}. Our method runs at 44 FPS in this 64-spp offline rendering application, showing a possible application to preview high-spp offline rendering.



\section{Analysis}

\subsection{\label{sec:ablation_study}\revisionFan{Modular Evaluations}}


In this subsection, we evaluate the effectiveness of our design choices. We modified the configuration of each compositing 
\revisionFan{module}
and compared the denoising quality with our complete architecture. Specifically, the complete architecture consists of a 6-layer convolutional neural network with \textit{RepVGG Block} and six fused filtering kernels . The kernel sizes range from $3$ to $13$. Other training settings are untouched in the ablation study.

\input{evaluate_RepVGG_Block}
\noindent \textbf{RepVGG Block.}
Practically, the denoising quality of our method is essentially influenced by the capacity of our \textit{ImportanceNet}, as can be seen in the quality comparison of \textit{Ours 6-layer} and \textit{Ours 3-layer} in Table~\ref{tab:BMFR_PSNR} and Table~\ref{tab:BMFR_SSIM}. However, the time budget of real-time applications has restricted the network architecture to be lightweight with a few narrow layers. As mentioned in Section~\ref{sec:prediction_phase}, the \textit{RepVGG Block} \cite{2021arXiv210103697D} utilizes the structural re-parameterization technique to convert one single branch CNN architecture to multiple branches with increased number of trainable parameters, and at the same time keeps the original inference speed. We present the error metrics of our 6-layer neural network
\revisionFan{and the KP method }
trained with and without \textit{RepVGG Block} in Table~\ref{tab:evaluate_RepVGG_Block}. In practice, \textit{RepVGG Block}'s over-parameterization property could avail the training process of small-sized networks, which brings us about 0.5dB improvement in PSNR for the presented test scenes. 
\revisionFan{
While the improvements to the KP method are limited due to its network parameter number is already large without the \textit{RepVGG Block} (nearly $3\times$ larger than ours).
}
Besides, we experimentally found that this structure has limited benefits for more shallow networks, specifically no improvement on average for the 2-layer networks of NBGD.

\input{KP_metrics}

\noindent \textbf{Kernel Construction and Kernel Prediction.}
To further compare the proposed method with the original kernel prediction method KPCN \cite{10.1145/3072959.3073708} under real-time application time budget, we design another two architecture variants of KPCN: a 2-layer network with filtering kernel size $13$ (KP-1) and a 5-layer network with filtering kernel size $7$ (KP-2). As presented in Table~\ref{tab:KP_metrics}, the comparison results show that there is a sharp decline in numerical quality if we reduce the network size of KP to satisfy the time budget (KP-1). In addition, the KP architecture with a proper network size and a small filtering kernel size has a similar execution speed with ours but produced a worse denoising quality (KP-2).

\begin{figure*}[htb]
  \centering
\includegraphics[width=1.0\linewidth]{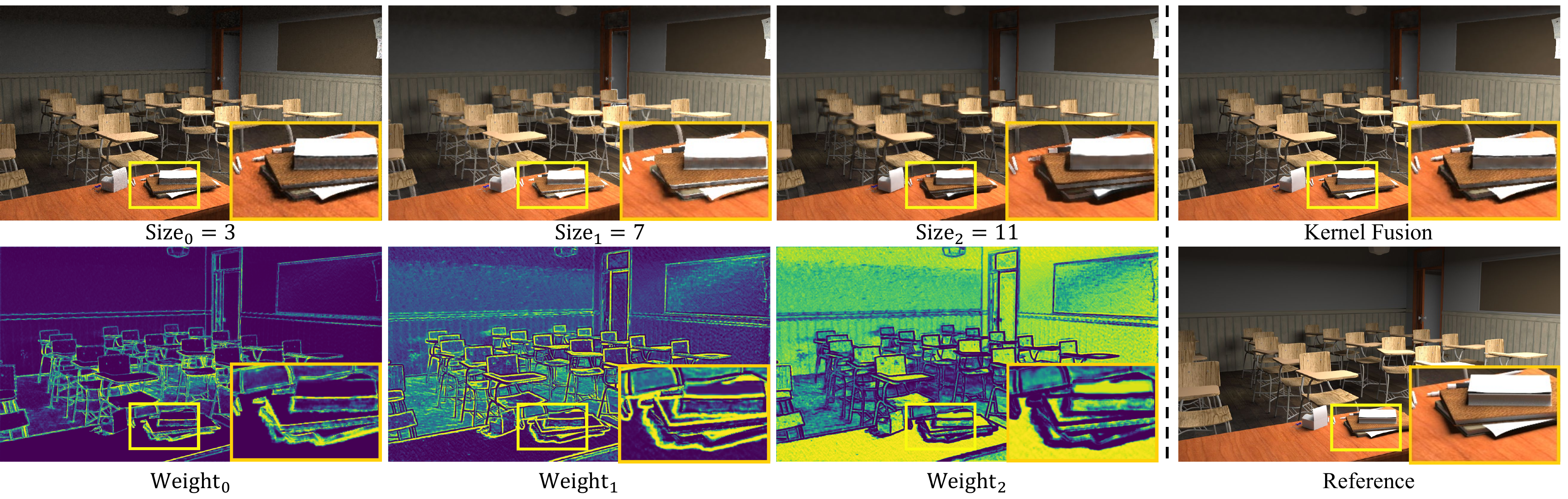}
  \caption{\label{fig:fusion}
           Visual quality comparisons between a kernel fusion architecture and the independently filtered results for different kernel sizes. Bright colors in weight map correspond to high averaging weight. The \textit{ImportanceNet} predicts higher averaging weight in high-frequency regions for small kernel filtered result while higher averaging weight in low-frequency region for large kernel filtered result, thus the fused result preserves the advantage of each sized filtering kernel.
           }
\end{figure*}

\noindent \textbf{Fusion of Different Sized Kernels.}
Most recent neural denoisers benefit from the design of denoising the input in different spatial resolutions. For example, NBGD constructs bilateral grids with different resolutions \cite{.20201133}, and the layer-based denoiser \cite{munkberg2020neural} splats samples to ordered layers then independently filter these layers. 
\revisionFan{
Likewise, we propose to denoise one pixel position with several varying-sized 
filtering kernels and combine the independently filtered results with a weighted average. Our kernel fusion module can overcomes the limitation of our correlated filtering kernels and bring the flexibility of explicitly handling varying-frequency noise as discussed in Section~\ref{sec:reconstruction_phase}. We visualize our kernel fusion scheme's prediction with kernel sizes $\{3, 7, 11\}$ in Figure~\ref{fig:fusion} to intuitively check its working behaviors. The end-to-end training manner makes our \textit{ImportanceNet} cooperate reasonably with the kernel fusion module. On the one hand, the filtering kernels use a cooperative manner to capture the signal changes from different directions. For example, around the book edges in Figure~\ref{fig:fusion}, the averaging weights of filtering kernels with size $3$ and $7$ have high values at opposite edge sides, which means they separately reconstruct pixels in different directions to contribute to the final denoised result when signal change happens. On the other hand, for high-frequency regions, our \textit{ImportanceNet} predicted higher averaging weight for small kernel filtered results and lower averaging weight for low-frequency regions. We also provide an interactive viewer in the supplementary to visualize the fusing kernel's weights and the corresponding averaging weights to examine reconstruction behavior better.
}

\revisionFan{

\begin{figure*}[htb]
  \centering
  \includegraphics[width=1.0\linewidth]{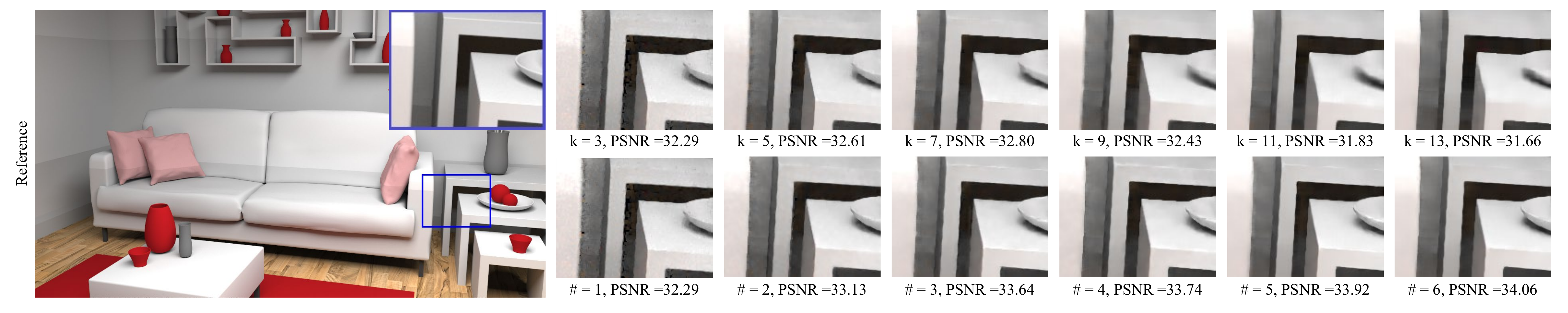}
  \caption{\label{fig:single_fusion}
            Comparison of the architecture with single filtering kernel and with fused filtering kernels on \textit{Living-room} test scene. The first inset row shows the reconstructed results with a single filtering kernel, and the $k$ means the size of constructed filtering kernel. The second inset row shows the reconstructed results with the kernel fusion module. The $\#$ stands for the fused filtering kernel count, and we cumulatively add filtering kernel from size $3$ to size $13$. The average PSNR is computed with the $60$ full-resolution frames.
           }
\end{figure*}

We also took the ablation study of our kernel fusion module to further evaluate its impact. As shown in Figure~\ref{fig:single_fusion}, we took two groups of experiments with the same network architecture. The first group (the first inset row in Figure~\ref{fig:single_fusion}) included six experiments trained without the kernel fusion module, which means they all constructed one single filtering kernel but with varying sizes ${k_i} = \{3, 5, 7, ..., 13\}$ to denoise. The second group (the section inset row in Figure~\ref{fig:single_fusion}) includes six experiments trained with the kernel fusion module. They have the increasing fusing kernel count ${\#_i}=\{1, 2, ..., 6\}$  and the kernel sizes are from $3$ with an incremental step of 2. For example, the third experiment has $3$ fused kernels with sizes $\{3, 5, 7\}$. The visual results and the average error metrics presented in Figure~\ref{fig:single_fusion} show that when using only one single filtering kernel, a small-sized filtering kernel can reconstruct sharper edges and a large-sized kernel performs better in low-frequency regions, while our kernel fusion module can take advantages of the varying-sized kernels with the weighted average. The average PSNR summarizes that the kernel fusion module outperforms any of its single compositing kernels in numerical error metrics. Besides, the first inset row also shows a failure case with leaking colors around the object edges when trained with only one single filtering kernel, and the second inset row shows that the leaking artifacts have decreased when trained with our kernel fusion module. 
}

\revisionFan{
Varying kernel sizes benefit filtering on different frequencies. In order to evaluate the impact, we also compared our method to the basic KP method modulated with the kernel fusion module, and to one variant of our method by fusing $6$ kernels with the same size $k_i=13$ in the supplementary document.
}

\begin{figure}[htb]
  \centering
  \includegraphics[width=1.0\linewidth]{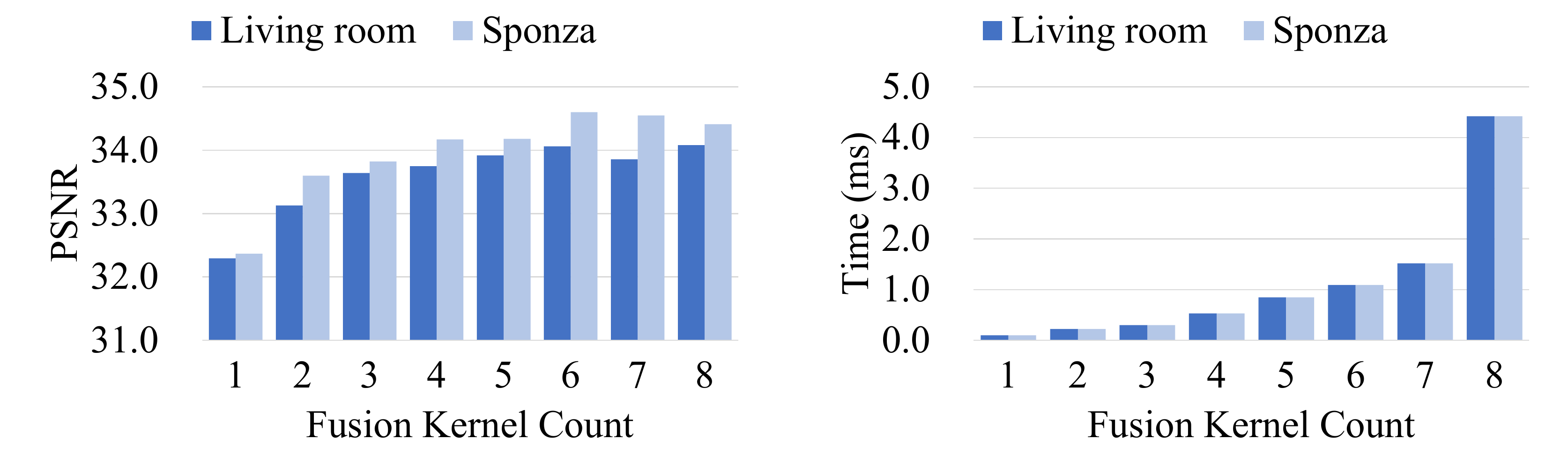}
  \caption{\label{fig:PSNR_time}
           Left: the denoising quality over different fusion kernel count, measured by average PSNR (higher is better) on the BMFR test scene. The kernel counts are cumulatively added with base size $k_b = 3$ and incremental step $k_s = 2$. Right: the corresponding execution time over fusion kernel count. Results show there are limited quality improvements but non-negligible performance reductions while fusing more than six kernels.
           }
\end{figure}

For each fusing kernel count, we also evaluated the average execution time of the corresponding kernel construction and fusion modules and present them in Figure~\ref{fig:PSNR_time}. It shows that increasing the fusing kernel count beyond six brings very limited quality improvement but significant time cost, so our implementation utilizes six filtering kernels, which is a good trade-off between denoising quality and execution speed.

\subsection{Limitations}

\noindent \textbf{Temporal Stability.}
Our method uses the widely adapted temporal accumulation preprocessing step to handle the temporal stability. Therefore, it has similar behaviors and limitations compared to other denoisers in terms of temporal stability. Please see the temporal stability measurement metric comparisons in the supplementary document. There exist some recent techniques worth trying in the future. For example, Xin et al. \cite{9194085} proposed to use a temporal loss between the warped outputs of the adjacent frames without introducing additional operations in their pipeline. 

\noindent \textbf{Large size kernel reconstruction.}
The PSNR in Figure~\ref{fig:single_fusion} experimentally shows that aggressively compressing a large-sized kernel, e.g., over a size of $9$,  in a single-channel importance map degenerates quality. Because the 
\revisionFan{compression}
ratio grows squarely over the kernel size $k$, our \textit{ImportanceNet} will predict an inadequate encoding for a large-size kernel and lead to a poorly behaved reconstructed result.  Besides, Figure~\ref{fig:PSNR_time} shows that when using the kernel fusion module, the quality improvement grows slowly if the kernel size is larger than $13$. This limitation constrains our approach from directly using very large kernels for offline applications (e.g., $21 \times 21$ in KPCN). However, it has little impact on the real-time application we aim for.

\noindent \textbf{Generalization to unseen effects.}
Similar to other neural denoisers, our method has a common generalization issue. While denoising specific rendering effects outside the training dataset, the method produces artifacts. This issue can be alleviated by enlarging the diversity of the training dataset.


\section{Conclusions}

We have presented our novel and practical weight sharing kernel prediction denoiser, which can denoise extreme low-spp Monte Carlo path traced images in real-time. At the core of our approach, we utilize an efficient neural network, \textit{ImportanceNet}, to learn to predict an encoding of the filtering kernel weights. Then we construct the filtering kernels  with a hand-crafted decoder in a splatting and fusing manner. The proposed weight sharing kernel prediction denoiser is scalable and allows us to tailor the design for different real-time applications. 
\revisionFan{
With this hand-crafted kernel constructor, our method can reduce nearly half the run-time cost and memory requirement of the basic kernel prediction method, and at the same time, produce comparable denoised results.
}

\bibliographystyle{eg-alpha-doi} 
\bibliography{egbibsample}

\end{document}

%% file: Tungsten_metrics.tex
\begin{table*}
  \centering
  \caption{
  Error metrics comparisons on 64-spp Tungsten test scenes of \textit{Kitchen} and \textit{Dining room}, averaged with 100 consecutive frames for each scene. We use the multi-resolution neural network architecture with two fused kernels each resolution.}
  \label{tab:Tungsten_metrics}
  \renewcommand\tabcolsep{4.0pt}
  \begin{tabular}{ccccccccccccc}
   \toprule
    \multirow{2}{*}{Scene}&
    \multicolumn{6}{c}{PSNR}&\multicolumn{6}{c}{SSIM}\\
    \cmidrule(lr){2-7} \cmidrule(lr){8-13}
                        &  NFOR  &  ONND  &  MR-KP &   KP   &  NBGD-7 &  Ours MR           &  NFOR  &  ONND  & MR-KP  &  KP  &  NBGD-7  &  Ours MR \\
    \midrule
    Kitchen             & 34.68  & 34.80  & \textbf{36.31}  & 36.03  & 35.53   & 35.88     & 0.973  & 0.973 & \textbf{0.978} & 0.977  & 0.974    & 0.976 \\
    Dining room         & 36.34  & 37.95  & 37.87  & \textbf{38.11}  & 35.98   & 38.03     & 0.980  & 0.970 & 0.980 & \textbf{0.982}  & 0.979    & 0.981 \\
    
  \bottomrule
\end{tabular}
\end{table*}

%% file: evaluate_RepVGG_Block.tex
\begin{table}
  \centering
  \caption{
            Error metrics comparisons for our \textit{ImportanceNet} and KP method trained with and without \textit{RepVGG Block}, averaged over 60 consecutive frames of each test scene.
          }
  \label{tab:evaluate_RepVGG_Block}
  \begin{tabular}{ccccc}
   \toprule
    \multirow{2}{*}{Scene}&
    \multicolumn{2}{c}{PSNR}&\multicolumn{2}{c}{SSIM}\\
    \cmidrule(lr){2-3} \cmidrule(lr){4-5}
                        & w/o                & w/                & w/o               & w/     \\
    \midrule
    Living room (Ours)  & 33.470             & \textbf{34.063}   & 0.9754            & \textbf{0.9789} \\
    Sponza (Ours)       & 33.925             & \textbf{34.600}   & 0.9813            & \textbf{0.9830} \\
    Living room (KP)    & 33.962             & \textbf{34.090}   & 0.9771            & \textbf{0.9781} \\
    Sponza (KP)         & 34.541             & \textbf{34.595}   & 0.9829            & \textbf{0.9830} \\
    
  \bottomrule
\end{tabular}
\end{table}

%% file: KP_metrics.tex
\begin{table}
  \centering
  \caption{
  PSNR and SSIM comparison averaged on all test scenes, and note that we use hold-out dataset partition, so this averaged metric is just for a rough analysis. Please see the detailed comparison of each test scene in our supplementary. KP-1 represents a KPCN variant with two layers CNN and filtering kernel size 13. KP-2 represents a KPCN variant with five layers CNN and filtering kernel size 7.
  }
  \label{tab:KP_metrics}
  \begin{tabular}{ccccc}
    \toprule
    Type              & KP-1   & KP-2   & Ours   \\
    \midrule
    PSNR              & 29.517  & 29.809  & 30.323  \\
    SSIM              & 0.941   & 0.946   & 0.951 \\
    Time (ms)         & 19.70  & 12.99  & 12.89  \\
  \bottomrule
\end{tabular}
\end{table}